\def\lesssim{\,\lower2truept\hbox{${<\atop\hbox{\raise4truept\hbox{$\sim$}}}$}\,}
\def\gtrsim{\,\lower2truept\hbox{${>\atop\hbox{\raise4truept\hbox{$\sim$}}}$}\,}

\documentclass{kapproc}
\usepackage[dvips]{graphicx}
\upperandlowercase
\kluwerbib

\begin{document}

\articletitle[A Physical Model for the joint evolution of QSOs and
Spheroids]{A Physical Model for the joint evolution of QSOs and
Spheroids}

\author{G.L. Granato,\altaffilmark{1} L. Silva,\altaffilmark{2}  G. De Zotti,\altaffilmark{1}
 A. Bressan\altaffilmark{1} and L. Danese\altaffilmark{3}}

\affil{ \altaffilmark{1}INAF, Padova, Italy, \
\altaffilmark{2}INAF, Trieste, Italy, \  \altaffilmark{3}SISSA,
Trieste, Italy}

\chaptitlerunninghead{Model for the Co-evolution of QSOs and of
their Spheroids} \anxx{Grabato\, G.L.}

\section{The model}

\begin{figure}[ht]
\centerline{\includegraphics[width=6.5truecm]{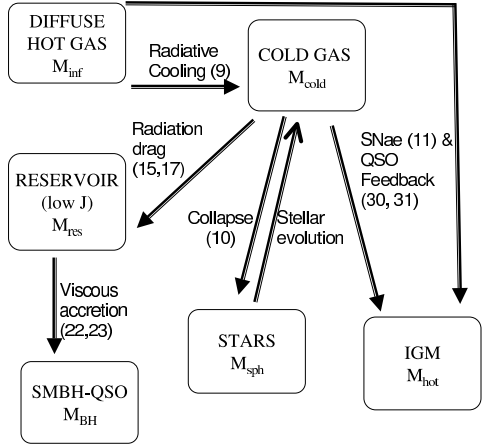}}
\caption{Scheme of the baryonic components included in the model
(boxes), and of the corresponding mass transfer processes
(arrows).} \label{schema}
\end{figure}

We present a detailed, physically grounded, model for the early
co-evolution of spheroidal galaxies and of active nuclei at their
centers. The model is based on very simple recipes, that can be
easily implemented. The main components and the transfer processes
accounted for are depicted in Fig.~\ref{schema}, while we defer to
Granato et al.\ (2004) for a full description. In summary, we
start from the diffuse gas within the dark matter halo falling
down into the star forming regions at a rate ruled by the dynamic
and the cooling times. Part of this gas condenses into stars, at a
rate again controlled by the local dynamic and cooling times. But
the gas also feels the feedback from supernovae and from active
nuclei, heating it and possibly expelling it from the potential
well. Also the radiation drag on the cold gas decreases its
angular momentum, causing an inflow into a reservoir around the
central black hole. Viscous drag then causes the gas to flow from
the reservoir into the black hole, increasing its mass and
powering the nuclear activity. Among the most novel point of our
work with respect to other semi-analytic approaches, we point out
the treatment of the evolution of QSO activity in the forming
spheroids, and of the resulting effects on galaxy evolution.

\section{Results}

In the shallower potential wells (corresponding to lower halo
masses and, for given mass, to lower virialization redshifts), the
supernova heating is increasingly effective in slowing down the
star formation and in driving gas outflows, resulting in an
increase of star/dark-matter ratio with increasing halo mass. As a
consequence, the star formation is faster within the most massive
halos, and the more so if they virialize at substantial redshifts.
Thus, in keeping with the proposition by Granato et al. (2001),
physical processes acting on baryons effectively reverse the order
of formation of galaxies compared to that of dark-matter halos.

A higher star-formation rate also implies a higher radiation drag,
resulting in a faster loss of angular momentum of the gas (Umemura
2001) and, consequently, in a faster inflow towards the central
black-hole. In turn, the kinetic energy carried by outflows driven
by active nuclei through line acceleration is proportional to
$L_{\rm QSO}^{3/2}$, and this mechanism can inject in the
interstellar medium a sufficient amount of energy to unbind it.
The time required to sweep out the interstellar medium, thus
halting both the star formation and the black-hole growth, is
again shorter for larger halos. For the most massive galaxies
($M_{\rm vir}\gtrsim 10^{12}\,M_\odot$) virializing at $3 \lesssim
z_{\rm vir} \lesssim 6$, this time is $< 1\,$Gyr, so that the bulk
of the star-formation may be completed before type Ia supernovae
can significantly increase the $Fe$ abundance of the interstellar
medium; this process can then account for the $\alpha$-enhancement
seen in the largest galaxies.

\begin{figure}[t]
\centerline{\includegraphics[width=6.5truecm]{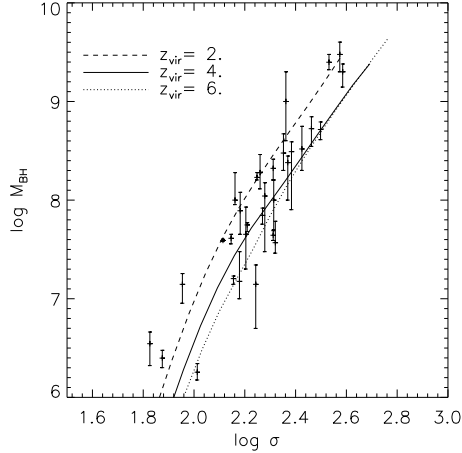}}
\caption{Predicted\ relationship between black-hole mass and
line-of-sight velocity dispersion of the host galaxy for different
virialization redshifts.} \label{MBHsigma}
\end{figure}

\begin{figure}
\centerline{\includegraphics[width=7truecm]{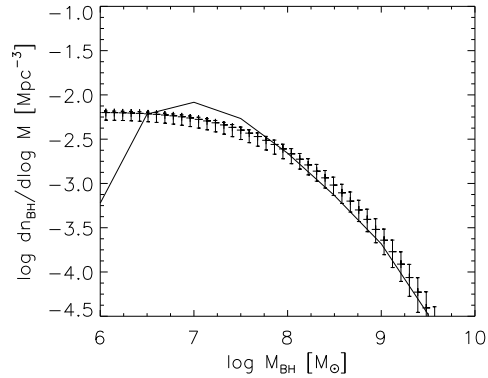}}
\caption{Predicted local black-hole mass function compared with
the recent estimate by Shankar et al. (in preparation). The total
mass density in BHs they derive is $\rho_{BH} \simeq 4 \times 10^5
M_{\odot} \mbox{Mpc}^{-3}$, 25\% less than our model. The decline
of the model at low $M_{\rm BH}$ is due to having considered only
objects with $M_{\rm vir} \geq 2.5\times 10^{11} M\odot$. }
\label{BHmassfunction}
\end{figure}

The interplay between star formation and nuclear activity
determines the relationship between the black-hole mass and the
mass, or velocity dispersion, of the host galaxy, as well as the
black-hole mass function. As illustrated by Figs.~\ref{MBHsigma}
and \ref{BHmassfunction}, the model predictions are in excellent
agreement with the observational data. A specific prediction of
the model is a substantial steepening of the $M_{\rm
BH}$--$\sigma$ relation for $\sigma \lesssim
150\,\hbox{km}\,\hbox{s}^{-1}$: the mass of the BH associated to
less massive halos is lower than expected from an extrapolation
from higher masses, because of the combined effect of SN heating,
which is increasingly effective with decreasing galaxy mass in
hindering the gas inflow towards the central BH, and of the
decreased radiation drag.

The ratio between QSO and SNae feedback is an increasing function
of the mass. At low mass the cumulative effect of QSO is almost
negligible with respect to that of SNae, but it becomes dominant
at intermediate mass (typically by a factor of a few) and high
mass (by a factor $\gtrsim 10$). Note that the QSO effect usually
increases exponentially with time, while that of SNae increases
more slowly. Thus the instantaneous QSO effect  becomes dominant,
if ever, only a few e-folding times before the maximum of QSO
activity.

\begin{figure}[ht]
\centerline{\includegraphics[width=6.5truecm]{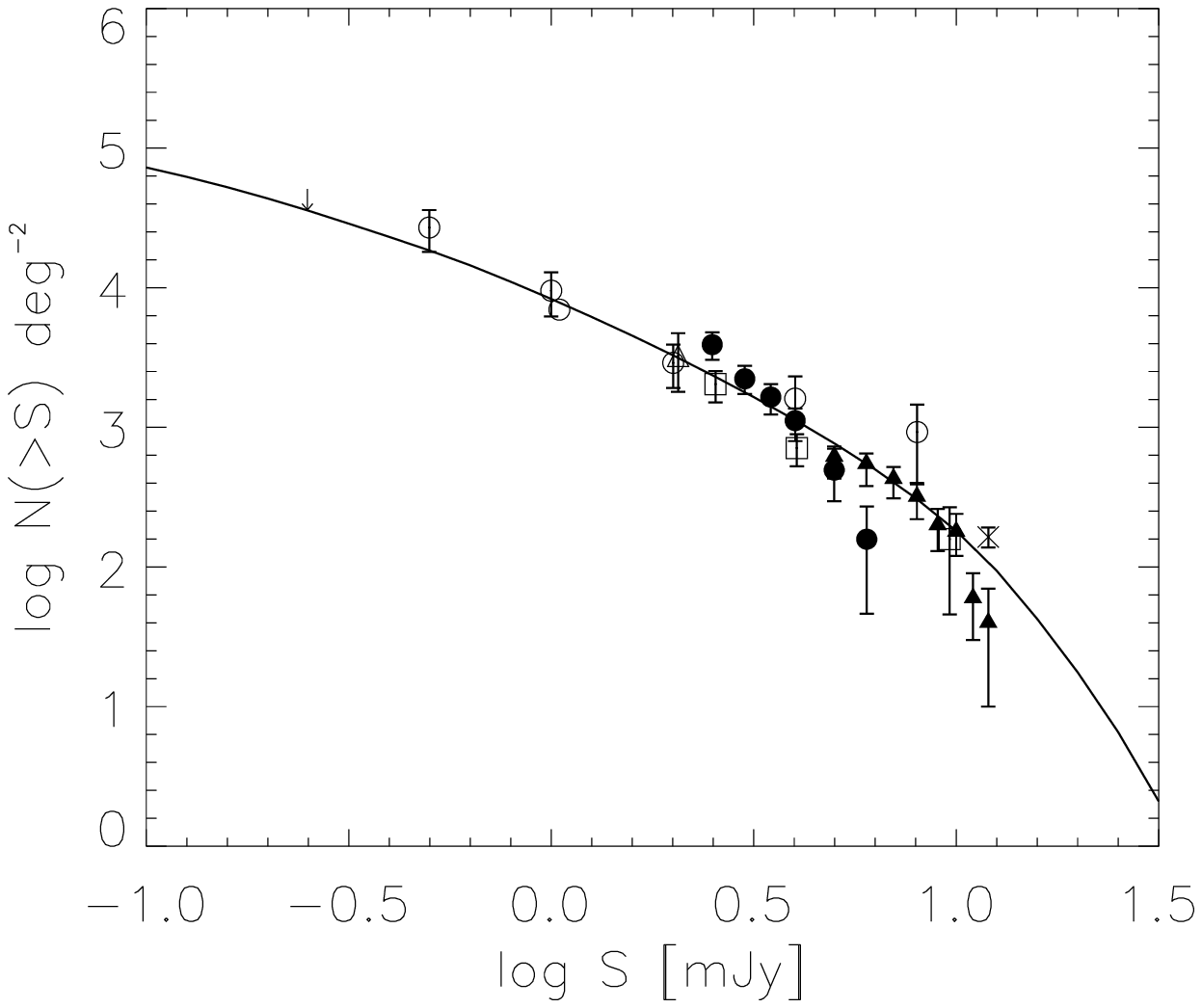}\includegraphics[width=6.5truecm]{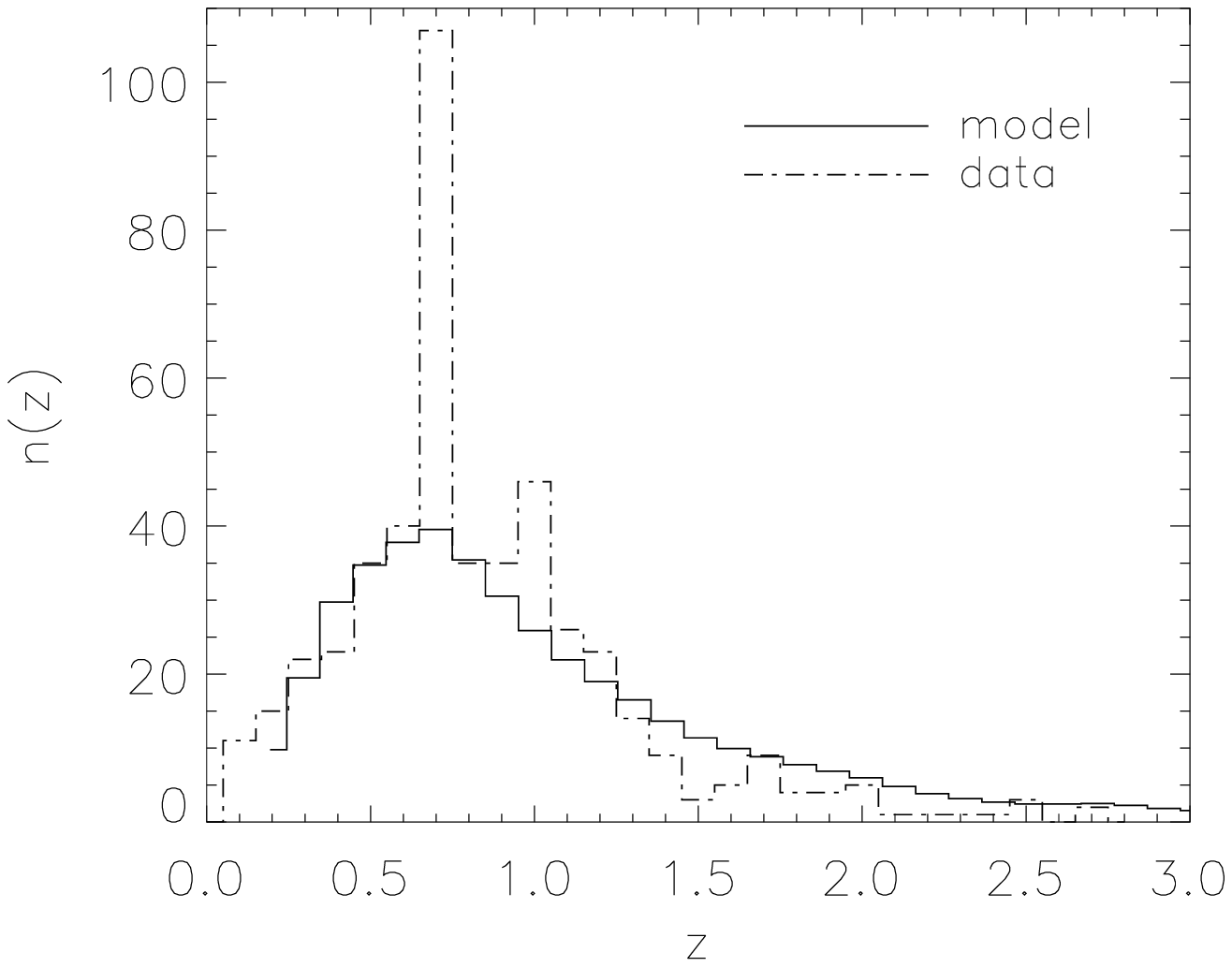}}
\caption{Left: predicted $850\,\mu$m extragalactic counts compared
with observed SCUBA counts. Right: predicted redshift distribution
of galaxies brighter than $K=20$ compared with the results of the
K20 survey}\label{figcount}
\end{figure}

Coupling the model with GRASIL (Silva et al. 1998), the code
computing in a self-consistent way the chemical and
spectrophotometric evolution of galaxies over a very wide
wavelength interval, we have obtained predictions for the sub-mm
counts and the corresponding redshift distributions as well as for
the redshift distributions of sources detected by deep K-band
surveys, which proved to be extremely challenging for all the
current semi-analytic models. The results, shown by
Fig.~\ref{figcount}, are again very encouraging.

The cosmic history of BH accretion, is in keeping with the common
notion that quasar activity peaks at redshift 2-3. Also, the
formation rate of spheroids derived in this paper is consistent
with that we computed in Granato et al.\ {2001} from a
deconvolution of the high-z luminosity function of QSO.

\begin{chapthebibliography}{1}

\bibitem[]{}
Granato, G.L., Silva, L., Monaco, P., Panuzzo, P., Salucci, P., De
Zotti, G., \& Danese, L. 2001, MNRAS, 324, 757

\bibitem[]{}
Granato, G.L.,  De Zotti, G., Silva,L., Bressan, A. \& Danese, L.
2004, ApJ, in press

\bibitem[]{}
Silva, L., Granato, G.L., Bressan, A., \& Danese, L. 1998, ApJ,
509, 103


\bibitem[{Umemura} 2001]{2001ApJ...560L..29U} Umemura,
M. 2001, ApJ,  560, L29


\end{chapthebibliography}
\end{document}